\documentclass[showpacs,preprintnumbers,amsmath,amssymb,amsfots]{revtex4}

\usepackage{amsmath,amsfonts,amssymb}
\usepackage{graphicx}
\usepackage{mathrsfs}

\newcommand{\beq}{\begin{equation}}
\newcommand{\eeq}{\end{equation}}
\newcommand{\beqa}{\begin{eqnarray}}
\newcommand{\eeqa}{\end{eqnarray}}
\def\nn{\nonumber\\}
\def\eq#1{(\ref{#1})}
\def\chr#1#2#3{\ensuremath{\Gamma^{#1}_{\,#2#3}}}
\def\cd#1{\ensuremath{\nabla_{#1}}}          
\def\pd#1{\ensuremath{\partial_{#1}}}        

\def\st{space-time}

\def\lab#1{\label{#1}}

\begin{document}


\title{{Infinite slabs and other weird plane symmetric space-times with constant positive density }}

\author{Ricardo E.\ Gamboa~Sarav\'{\i}}

\email{quique@fisica.unlp.edu.ar} \affiliation{Departamento de
F\'\i sica, Facultad de Ciencias Exactas, Universidad Nacional de
La Plata,\\and IFLP,
CONICET.\\C.C. 67, 1900 La Plata, Argentina}%


\begin{abstract}

We present  the exact  solution of Einstein's equation corresponding to a static and plane symmetric distribution of matter with constant positive density located below $z=0$. This solution depends essentially on two constants: the density $\rho$ and a parameter $\kappa$. We show that this space-time finishes down below at an inner singularity at finite depth. We match this solution to the vacuum one and compute the external gravitational field in terms of  slab's parameters. Depending on the value of $\kappa$, these slabs can be attractive, repulsive or neutral. In the first case, the space-time also finishes up above at another singularity.  In  the other cases,  they turn out to be semi-infinite and  asymptotically flat when $z\rightarrow\infty$.

We also find solutions consisting of  joining   an attractive slab and a repulsive one, and two neutral ones. We also discuss how to assemble a ``gravitational capacitor" by inserting a slice of vacuum between two such slabs.

\end{abstract}

\maketitle


\section{Introduction}

Due to  the complexity of Einstein's field equations, one
cannot find exact solutions except in spaces of rather high
symmetry, but very often with no direct physical application.
Nevertheless, exact solutions can give an idea of the qualitative
features that could arise in General Relativity,  and so, of possible
properties of realistic solutions of the field equations.

In this paper we want to illustrate some curious features of gravitation by means of a simple solution:  the gravitational field of a static plane symmetric relativistic perfect incompressible  fluid with positive density located below $z=0$.
Because of the symmetry required,  the  exterior   gravitational field turns out to be Taub's plane vacuum solution \cite{taub}. The internal solution was also found by  Taub \cite{taub2}.

Here we  match  both solutions, this corresponds to the plane symmetric counterpart of the Schwarzschild solution for a sphere of constant density \cite{schw}.

The solutions turn out to be attractive, repulsive or neutral
depending on the value of a parameter. These  space-times  present
some  somehow astonishing  properties without counterpart in
Newtonian gravitation:

Attractive solutions finish high above at an empty (free of matter) repelling boundary  where space-time curvature diverges. These  singularities are not the sources of the fields, but they arise owing to  the attraction of distant matter  as pointed out in \cite{gs}.

Repulsive slabs explicitly show how  negative but finite pressure can dominate the attraction of the matter.

We also consider  matching   two different internal solutions.

In Sec. II we present a simple and complete  derivation  of Taub's
interior solution. In Sec. III we make a detailed study of it. In
Sec. IV we discuss how solutions can be matched.

Throughout this paper, we adopt the convention in which the \st\
metric has signature $(-\ +\ +\ +)$, the  system of units in which
the speed of light $c=1$,  Newton's gravitational constant $ G=1$
and $g$  denotes gravitational field and not the determinant of
the metric.

\section{The Taub interior solution}

 In this section we consider  the solution  of Einstein's
equation corresponding to  a static and plane symmetric
distribution of matter with constant positive density and plane
symmetry. That is, it must be invariant under translations in the
plane and under rotations around its normal. The matter we shall
consider is a perfect fluid satisfying the equation of state $
\rho= \rho_0$, where $\rho_0$ is an arbitrary positive constant.
The stress-energy tensor is \beq T_{ab}= (\rho+p)\,u_au_b+p\,
g_{ab}\, ,\eeq where $u^a$ is the velocity of fluid elements.

Due to the plane symmetry and staticity, following \cite{taub} we
can find coordinates $(t, x, y, z)$ such that \beq \lab {met}
ds^2= - \mathcal{G}(z)^{2 }\ dt^2+ e^{2V(z)}\left(dx^2+dy^2
\right)+dz^2\,.\eeq%
That is, it is the more general metric admitting the Killing vectors $\pd x$, $\pd y$, $x\pd y-y\pd x$ and $\pd t$.

The non identically vanishing components of the Einstein tensor are
\beqa \lab {gtt} G_{tt}=-\,\mathcal{G}^2 \left(2\,V''+3\, V'^2\right)\\
\lab {gii} G_{xx}=G_{yy}= e^{2V} \left({\mathcal{G}''}/{\mathcal{G}}+{\mathcal{G}'}/{\mathcal{G}}\;V'+V''+V'^2\right)\,,\\
\lab {gzz} G_{zz}=  V' \left(2\ {\mathcal{G}'}/{\mathcal{G}}+  V' \right) ,
 \eeqa where a prime $(')$ denotes differentiation with respect to $z$.%

On the other hand, since the fluid must be static,
$u_a=(\mathcal{G},0,0,0)$,  so \beq T_{ab}= \text{diag}\left(\rho\,
\mathcal{G}^{2},p\, e^{2V},p\, e^{2V},p\right)\,, \eeq where
 $p$  depends only on the z-coordinante.
 Thus, Einstein's equations, i.e., $G_{ab}=8 \pi T_{ab}$, are
\beqa \lab {gtt1} 2\,V''+3\, V'^2= -8 \pi{\rho}\,, \\%
\lab {gii1}{\mathcal{G}''}/{\mathcal{G}}+{\mathcal{G}'}/{\mathcal{G}}\;V'+V''+V'^2 =
8 \pi{p}\, ,\\
\lab {gzz1}  V' \left(2\ \mathcal{G}'/{\mathcal{G}}+  V' \right) =
8 \pi{p}\,.
 \eeqa%

Moreover, $\cd a T^{ab}=0$ yields
\beq \lab{ppr} p' = -(\rho+p)\,\mathcal{G}'/\mathcal{G}\,.\eeq Of course, due to Bianchi's identities equations,
(\ref{gtt1}), (\ref{gii1}), (\ref{gzz1}) and (\ref{ppr}) are not
independent, so we shall here use  only \eq{gtt1}, \eq{gzz1}, and
\eq{ppr}.

Since $\rho$ is constant, from \eq {ppr} we readily find
 \beq\lab{p}p  = C_p /\mathcal{G}(z)-\rho,  \eeq where $C_p$ is an arbitrary constant.

By setting 
$W(z)= e^{{3}V(z)/{2}}$, 
we can write  \eq{gtt1} as $W''=-{{6\pi\rho}}\, W$, and its
general solution  can be written as \beqa \lab{W1}W(z)= C_1\,
\sin{(\sqrt{{6\pi\rho}}\ z+C_2)},\, \eeqa
where $C_1$ and $C_2$ are arbitrary constants. 
Therefore,  we have
 \beq \lab{V} V(z)=\ln\left(
{C_1\,\sin{(\sqrt{{6\pi\rho}}\ z+C_2)}}\right)^{\frac{2}{3}}.\eeq

 Now, by replacing \eq {p} into \eq {gzz1} we get the  first order linear differential equation which $\mathcal{G}(z)$ obeys
\beqa  \lab {gzz2} \mathcal{ G}'=-\left(\frac{4\pi\rho}{V'}+\frac{V'}{2}\right)\mathcal{G} + \frac{4\pi C_p}{V'}\ \\ =-{\sqrt{{6\pi\rho}}}\left(\tan u+\frac{1}{3}\cot u\right)\mathcal{G} +{\sqrt{{6\pi\rho}}}\,\ \frac{ C_p}{\rho}\tan u \,,\eeqa
where $u=\sqrt{{6\pi\rho}}\ z+C_2$, and in the last step we have made use of \eq {V}.
The general solution of \eq {gzz2} can be written as%
\beqa  \mathcal{G}=  \frac{\cos u}{(\sin u)^{1/3}}\left(\text{const.} \ +\frac{C_p}{\rho}\int\frac{(\sin u)^{4/3}}{(\cos u)^2}du\right)\nn \lab {G} = C_3\, \frac{\cos u}{\left(\sin u \right)^{1/3}}+\frac{3 C_p}{7\rho}\  {\sin^2\! u}\,\, _2F_1\!\Bigl(1,\frac{2}{3};\frac{13}{6};\sin^2u \Bigr), \eeqa %
where $C_3$ is another arbitrary constant, and $_2F_1(a,b;c;z)$ is the Gauss hypergeometric function (see for example \cite{tablarusa}).

Therefore,  the line element \eq{met} becomes%
 \beqa \lab{met1}
ds^2=  -  \mathcal{G}(z)^{2}\,
 dt^2 + \left({C_1\,\sin u}\right)^{\frac{4}{3}}\left(dx^2+dy^2\right) + dz^2,
 \eeqa%
 where $\mathcal{G}(z)$ is given in \eq {G} and  $u=\sqrt{{6\pi\rho}}\ z+C_2$.
Thus, the solution contains five arbitrary constants: $\rho$, $C_p$, $C_1$, $C_2$, and $C_3$. The range of the coordinate $z$ depends on the value of these constants.

For  $\rho>0$, $C_p>0$  and $C_3>0$, this solution was found by
Taub  \cite{taub2}. When $C_p =0$, it is clear from \eq{p} that
$p(z)=-\rho$, and the solution \eq{met1} turns out to  be a vacuum
solution with a cosmological constant $\Lambda=8\pi\rho$
\cite{NH}. For   $C_p=0$ and $\rho\rightarrow0$, by an appropriate
choice of the constants we can readily see that \eq {met1} becomes
\beqa \lab{taub} ds^2=- (1-3gz)^{-\frac{2}{3}}\,
 dt^2 + (1-3gz)^{\frac{4}{3}}\left(dx^2+dy^2\right) +
 dz^2,\nn\nn
-\infty<t<\infty,\quad-\infty<x<\infty,\quad-\infty<y<\infty,\quad0<1-3gz<\infty \,,  \eeqa%
where $g$ is an arbitrary constant.  In \eq{taub} the coordinates
have been chosen
 in such a way that it describes a homogeneous  gravitational
 field $g$ pointing in the negative $z$-direction in a neighborhood of  $z=0$. The metric \eq{taub} is Taubs's vacuum plane solution \cite{taub} in the coordinates
 used in \cite{gs}.

Nevertheless, the solution \eq{met1} has a  wider range of validity.
 For {\em abnormal } matter some interesting
solutions also arise, but  the complete analysis turns out to be
somehow involved. So, for the sake of clarity, we shall confine
our attention to positive values of $\rho$ and $C_p \neq0$,
leaving the complete study to a forthcoming publication
\cite{gs2}.

Notice  that the metric \eq{met1} has a \st\, curvature
singularity when $u=0$, since straightforward computation of the
scalar quadratic in the
Riemann tensor yields%
\beqa\lab{RR} R_{abcd}R^{abcd}=4
\left({\mathcal{G}''^2}+2\,{\mathcal{G}'^2}\,V'^2\right)/{\mathcal{G}^2}+4\left(2\,V''^2+4\,V''V'^2+3\,V'^4\right)\nn
  =\frac{256}{3}\,\,\pi^2\rho^2\,\left(2+{\sin^{-4} u}+ \frac{3}{4} \left(\frac{p}{\rho}+1\right)\left(\frac{3p}{\rho}-1\right)\right), \eeqa
so $ R_{abcd}R^{abcd}\rightarrow\infty$ when $ u\rightarrow0$.

\section{The properties of the function $\mathcal{G}(z)$}


In this section we shall study in detail the properties of the solution for the case $\rho>0$ and $C_p \neq0$.

Now, it  is clear from
(\ref{gtt1}), (\ref{gii1}), (\ref{gzz1}) and (\ref{ppr})
 that field equations are invariant under the transformation
  $z\rightarrow \pm z+z_0$, i.e., z-translations and mirror reflections
across any plane $z\!=$const. Thus, if
$\{\mathcal{G}(z),V(z),p(z)\}$ is a solution  $\{\mathcal{G}(\pm
z+z_0),V(\pm z+z_0),p(\pm z+z_0)\}$ is another one, where $z_0$ is
an arbitrary constant. Therefore, taking into account that
${u=\sqrt{{6\pi\rho}}\,z+C_2}$, without loss of generality the
consideration of the case  $0<u<\pi/2$  shall suffice.

By an appropriate rescaling of the coordinates
$\left\{t,x,y\right\}$, without loss of generality,   we can write
the metric  \eq {met1} as \beqa \lab{met2} ds^2=  -
\mathcal{G}(z)^{2}\, dt^2 + \,\sin^{\frac{4}{3}}
u\,\left(dx^2+dy^2\right) + dz^2,\nn
-\infty<t<\infty,\quad-\infty<x<\infty,\quad-\infty<y<\infty,\quad0<u=\sqrt{{6\pi\rho}}\ z+C_2\leq\pi/2, \eeqa%
and \eq{G} as \beqa \lab{G8} \mathcal{G}(z)=  \frac{\kappa C_p}{ \rho}\, \frac{\cos u}{\sin^{1/3} u }+\frac{3   C_p}{7 \rho}\  {\sin^{2} u}\,\,\, _2F_1\!\Bigl(1,\frac{2}{3};\frac{13}{6};\sin^2u \Bigr), \eeqa %
where $\kappa$ is an arbitrary constant.

 By replacing \eq{G8} into \eq {p}, we see that
the pressure is independent of $C_p$. On the other hand, since
$\mathcal{G}(z)$ appears squared in the metric, it suffices to
consider $C_p>0$. Furthermore, rescaling again the coordinate $t$,
we may set $C_p=\rho$. Thus, \eq {G8} becomes
\beqa \lab{G2} \mathcal{G}(z) =G_\kappa(u)= \kappa \, \frac{\cos u}{\sin^{1/3} u }+
\frac{3  }{7 }\  {\sin^{2} u}\,\,\, _2F_1\!\Bigl(1,\frac{2}{3};\frac{13}{6};\sin^2u \Bigr), \eeqa %
where $G_\kappa(u)$ was defined for future use, and we recall that
${u=\sqrt{{6\pi\rho}}\,z+C_2}$.  Furthermore, \eq {p} becomes
\beq\lab{p3}p(z)  = \rho \left(1 /\mathcal{G}(z)-1\right). \eeq

Therefore the solution depends on two essential parameters, $\rho$ and $\kappa$.
We shall discuss in detail the properties of the functions $\mathcal{G}(z)$ and $p(z)$
depending on the value of the constant $\kappa$.

Now, the hypergeometric function in the last equation  is a
monotonically increasing continuous positive function of $u$ for
$0\leq u\leq \pi/2$, since $c-a-b={1}/{2}>0$. Furthermore
\cite{abra} \beqa
_2F_1\!\Bigl(1,\frac{2}{3};\frac{13}{6};0\Bigr)=1,\,\,\,\,\text{and}\,\,\,\,_2F_1\!\Bigl(1,\frac{2}{3};\frac{13}{6};1\Bigr)=
\frac{7}{3}. \eeqa Therefore, we readily see from \eq {G2} that,
no matter what the value of $\kappa$ is,
 $\mathcal{G}(z)|_{u=\pi/2}=1$ and we get then from \eq {p3} that
$p(z)$  vanishes at $u=\pi/2$.

On the other hand, since
\beq \lab{G0}\mathcal{G}(u)= \kappa\, u^{-1/3}+O(u^{5/3})\,\,\,\,\,\,\,\,\, \text{as}\,\,\,\,\,\, u\rightarrow0\,, \eeq
 $\mathcal{G}(z)|_{u=0}=0$ if $\kappa=0$, whereas  it diverges if $\kappa\neq0$.

For the sake of clarity, we shall analyze separately the cases $\kappa>0$, $\kappa=0$, and $\kappa<0$.

\subsubsection{$\kappa>0$}

\begin{figure}
\begin{center}
\includegraphics[height=12cm]{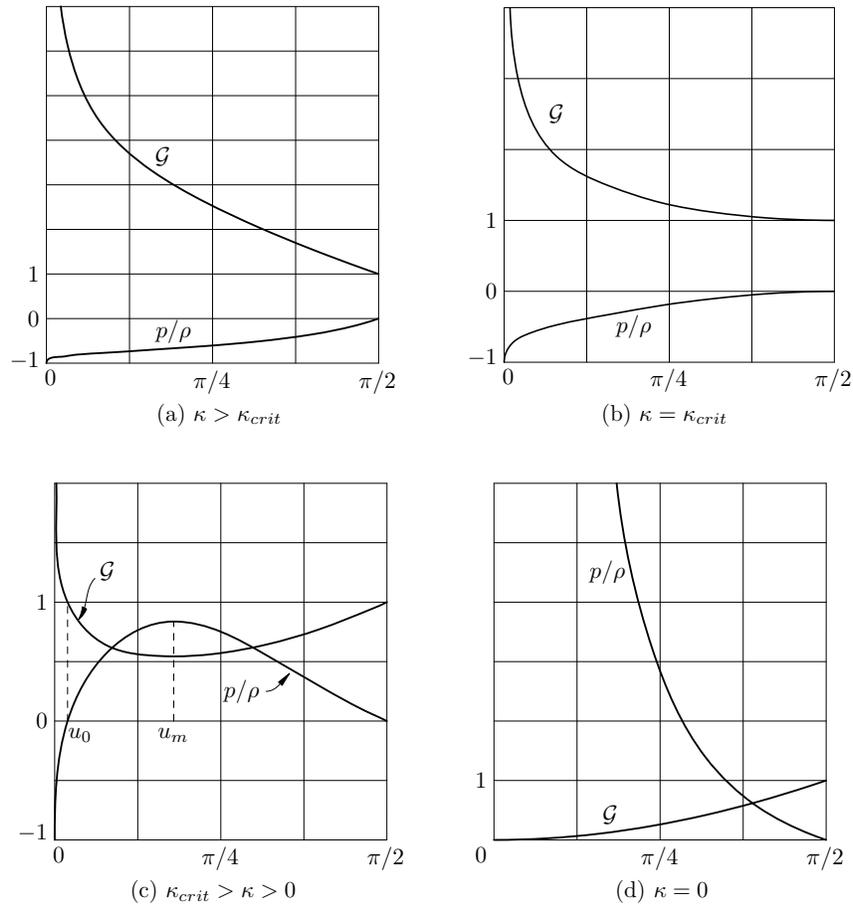}
\caption{\label{Gp}$\mathcal{G}(z)$ and $p(z)$, as functions of
$u$, for decreasing values of $\kappa\geq0$.}\end{center}
\end{figure}

In this case, it is clear from \eq {G2} that  $\mathcal{G}(z)$ is
positive definite when $0<u\leq\pi/2$. On the other hand, from \eq
{gii1} and \eq {gzz1} we get \beqa \lab {G''}
\mathcal{G}''=\mathcal{G}' V'-\mathcal{G} V''=
-\Bigl(V''+\frac{V'^2}{2}+4\pi\rho\Bigr)\mathcal{G}+4\pi C_p  \nn
= V'^2 \mathcal{G}+4\pi C_p, \eeqa%
where we have made use of \eq {gzz2} and \eq {gtt1}. Then also
$\mathcal{G}''$ is  positive definite in $0<u\leq\pi/2$, and so
$\mathcal{G}'$ is a monotonically increasing continuous function
of $u$ in this interval.

Now, taking into account that $\,\mathcal{G'}=\pd z \mathcal{G}=
\sqrt{6\pi\rho}\ \pd u\mathcal{G}$, a straightforward computation from \eq {G2} shows that
\beqa
\mathcal{G}'(z)\lab{G'0}= -\frac{\kappa \sqrt{6\pi\rho} }{3}u^{-4/3}+O(u^{2/3})\,\,\,\,\,\,\,\,\,
\text{as}\,\,\, u\rightarrow0, \eeqa and
\beqa \lab{G'1}\mathcal{G}'(z)|_{u=\pi/2}=\sqrt{6\pi\rho}\left(\kappa_{crit}-\kappa\right)\,,
 \eeqa
where \beq \lab{kapa}
\kappa_{crit}=\sqrt{\pi}\,{\Gamma(7/6)}/{\Gamma(2/3)}=1.2143\dots\eeq

If $\kappa\geq \kappa_{crit}$, $\mathcal{G}'$ is negative for
small enough values of $u$ and non-positive  at ${u=\pi/2}$, hence
$\mathcal{G}'$ is negative in $0<u<\pi/2$, so $\mathcal{G}(z)$ is
decreasing, and then $\mathcal{G}(z)>\mathcal{G}(z)|_{u=\pi/2}=1$
in this interval (see Fig.\ref{Gp}(a) and Fig.\ref{Gp}(b)).

For $\kappa_{crit}>\kappa>0$, $\mathcal{G}'$ is negative for
sufficiently small values of $u$ and positive at $\pi/2$. So there
is one   (and only one) value  $u_{m}$ where it vanishes. Clearly
$\mathcal{G}(z)$ attains a local minimum  there. Hence, there is
one (and only one) value  $u_0$ ($0<u_0<\pi/2$) such that
$\mathcal{G}(z)|_{u=u_0}=\mathcal{G}(z)|_{u=\pi/2}=1$, and then
$\mathcal{G}(z)<1$ when $u_0<u<\pi/2$ (see Fig.\ref{Gp}(c)).

Since $\mathcal{G}(z)>0$, it is clear from \eq {p3} that $p(z)>0$
if $\mathcal{G}(z)<1$, and $p(z)$  reaches a maximum when
$\mathcal{G}(z)$ attains a minimum.

Therefore, for $\kappa\geq \kappa_{crit}$,  $p(z)$ is  negative
when $0\leq u<\pi/2$ and increases monotonically from $-\rho$ to
$0$ (see Fig.\ref{Gp}(a) and  Fig.\ref{Gp}(b)).

On the other hand, for $\kappa_{crit}>\kappa>0$, $p(z)$ grows from
$-\rho$ to a maximum positive value when $u=u_{m}$ where it starts
to decrease and vanishes at $u=\pi/2$. Thus, $p(z)$ is negative
when $0<u<u_0$ and positive when $u_0<u<\pi/2$ (see
Fig.\ref{Gp}(c)).

\subsubsection{$\kappa=0$}

In this case, it is clear from \eq {G2} that  $\mathcal{G}$  monotonically increases
with $u$ from $0$ to $\mathcal{G}(z)|_{u=\pi/2}=1$. Therefore, $p$ is  a
monotonically decreasing positive continuous function of $u$ in $0<u<\pi/2$
(see Fig.\ref{Gp}(d)). Furthermore, at $u=0$ it diverges, since
\beqa
p(z)\sim \frac{7\rho }{3}u^{-2}\rightarrow+\infty\,\,\,\,\,\,\,\,\, \text{as}\,\,\, u\rightarrow0. \eeqa

\subsubsection{$\kappa<0$}
\begin{figure}[h]
\begin{center}
\includegraphics[height=6cm]{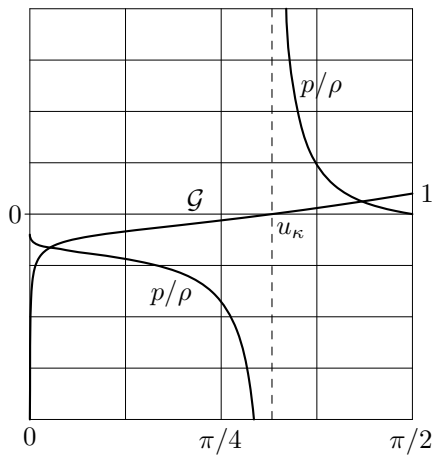}
\caption{\label{Gpneg}$\mathcal{G}(z)$ and $p(z)$ as functions of
$u$ for $\kappa<0$.}\end{center}\end{figure}

In this case, we see from \eq{G'0} that $\mathcal{G}'$ is positive
when $u$ takes  small enough values, and from \eq{G'1} that it is
also positive when $u$ is near to $\pi/2$.

Now, suppose that $\mathcal{G}'(z)$ attains a local minimum when
$u=u_1$ ($0<u_1<\pi/2$), then $\mathcal{G}''(z)|_{u=u_1}=0$.
Hence, we get from \eq{G''} that  $\mathcal{G}(z)|_{u=u_1}<0$, and
taking into account that
$V'(z)|_{u=u_1}={2\sqrt{6\pi\rho}}/{3}\cot u_1>0$, we see from
\eq{gzz2}  that  $\mathcal{G}'(z)|_{u=u_1}>0$. Thus, we have shown
that $\mathcal{G}'(z)$ is a continuous positive definite function
when $0<u\leq\pi/2$ if $\kappa<0$.

Therefore, in this case, $\mathcal{G}(z)$ is a continuous function
monotonically increasing  with $u$ when $0<u\leq\pi/2$. Since it
is negative for  sufficiently small values of $u$ and  $1$ when
$u=\pi/2$ it must vanish at a unique value of $z$ when
$u=u_\kappa$ (say). Furthermore $\mathcal{G}(z)<1$ when
$0<u<\pi/2$.

Clearly, we get from \eq {G2} that $u_\kappa$ is given implicitly
in terms of $\kappa$ through \beq \lab{k}{\kappa }=  -\frac{3}{7}\
\frac{{\sin^{7/3} u_\kappa}}{\cos u_\kappa}\,\,
_2F_1\!\Bigl(1,\frac{2}{3};\frac{13}{6};\sin^2u_\kappa
\Bigr)\,.\eeq We can readily see from \eq{k} that $u_\kappa$ is a
monotonically decreasing function of $\kappa$ in
$-\infty<\kappa<0$, and it tends to $\pi/2$ when
$\kappa\rightarrow-\infty$ and to $0$ when $\kappa\rightarrow
0^{-}$.

From \eq {p3}, it is clear that $p(z)$ diverges when $u=u_\kappa$.
Furthermore, \eq {p3} also shows  that $p(z)<0$ when
$\mathcal{G}(z)<0$, and taking into account that
$\mathcal{G}(z)<1$, that $p(z)>0$ when $\mathcal{G}(z)>0$.
Therefore, $p(z)$ is negative when $0<u<u_\kappa$ and positive
when $u_\kappa<u<\pi/2$ (see Fig.\ref{Gpneg}).

On the other hand, we see from \eq{RR} that, when $\kappa$ is
negative, another \st\, curvature singularity arises at $u_\kappa$
(besides  the one at $u=0$) since $p$ diverges there.

Therefore, if $\kappa$ is negative, the metric \eq{met2} describes two very
different space-times:

(a) For $0<u<u_\kappa$, the whole  space-time is trapped between
two singularities separated by a finite distance
$\sqrt{6\pi\rho}\,u_\kappa$. This is a space-time full of a
fluid with constant positive density $\rho$ and negative pressure
$p$ monotonically decreasing with $u$,  and $p(z)|_{u=0}=-\rho$
and $p(z)\rightarrow-\infty$ as $u\rightarrow u_\kappa$.

(b) For $u_\kappa<u<\pi/2$, the   pressure is positive and
monotonically decreasing with $u$, $p(z)\rightarrow \infty$ as
$u\rightarrow u_\kappa$ and $p(z)|_{u=\pi/2}=0$.

\section{Matching solutions}

In this section, we shall discuss  matching  the interior
solution to the vacuum one, as well as  joining  two
interior solutions facing each other.

Since the equations are invariant under $z$-translation, we can
choose to match the solutions at $z=0$ without losing  generality.
So we select $C_2=\pi/2$, and then  \eq {G2} becomes \beqa
\lab{G3} \mathcal{G}(z)=G_\kappa(\sqrt{6\pi\rho}\,z+\pi/2)= -
\kappa \, \frac{\sin (\sqrt{6\pi\rho}\,z)}{\cos^{1/3}
(\sqrt{6\pi\rho}\,z )} +\frac{3  }{7 }\  {\cos^{2}
(\sqrt{6\pi\rho}\,z)}\,\,\, _2F_1\!\Bigl(
1,\frac{2}{3};\frac{13}{6};\cos^2(\sqrt{6\pi\rho}\,z) \Bigr)\,. \eeqa %
Therefore, the metric \eq{met2} reads \beqa \lab{met3}ds^2=  -
G_\kappa(\sqrt{6\pi\rho}\,z+\pi/2)^{2}\, dt^2 +
\,\cos^{\frac{4}{3}} (\sqrt{6\pi\rho}\,z)\,\left(dx^2+dy^2\right)
+ dz^2,\nn
-\infty<t<\infty,\quad-\infty<x<\infty,\quad-\infty<y<\infty,\quad{-\sqrt{\pi/24\rho}}<z\leq0\,. \eeqa%
Notice that  it holds that $g_{tt}(0)=-\mathcal{G}(0)^2=-1$,
$g_{xx}(0)=g_{yy}(0)=1$, and from \eq{p3} we get $p(0)=0$.

Furthermore, we shall impose the continuity of $\pd
zg_{tt}$, and from \eq{G'1} we have \beqa \lab{G'2} \pd
zg_{tt}(0)|_{interior}=
-2\,\mathcal{G}(0)\mathcal{G}'(0)=-2\sqrt{6\pi\rho}\left(\kappa_{crit}-\kappa\right)\,.
 \eeqa

\subsection{Matching solutions to vacuum solutions}

Here, we shall consider the gravitational field of a planar
symmetric distribution of matter with constant density $\rho$
sitting below $z=0$. This is the plane symmetric counterpart of
Schwarzschild's solution for a sphere of incompressible fluid
\cite{schw}.

 In this case, we see from \eq{taub} that the corresponding
exterior solution for $z>0$ is \beqa \lab{met5} ds^2=-
(1-3gz)^{-{2}/{3}}\,
 dt^2 + (1-3gz)^{{4}/{3}}\left(dx^2+dy^2\right) +
 dz^2,\nn\nn
-\infty<t<\infty,\quad-\infty<x<\infty,\quad-\infty<y<\infty,\quad0\leq z<\begin{cases}1/3g
&\text{if}\,\,\, g>0\\\infty&\text{if}\,\,\,g\leq 0 \end{cases}\,,  \eeqa%
which for $g>0$ is the  region finishing at the singularity of
Taubs's vacuum plane solution,  whereas it is the asymptotic flat
tail of it for $g<0$. It describes a homogeneous gravitacional
field $-g$ in the vertical (i.e., $z$) direction \cite{gs}.

Since $ g_{tt}(0)|_{exterior}=-1$, $ g_{xx}(0)|_{exterior}=g_{yy}(0)|_{exterior}=1$
the continuity of the metric is assured,  and concerning the derivative we have
\beqa \lab{G'3} \pd zg_{tt}(z)|_{exterior}= -2g\,
(1-3gz)^{-{5}/{3}}\,.
 \eeqa
Therefore,   by comparing with \eq{G'2}, we see that the
continuity of  $\pd zg_{tt}$ yields \beqa
\lab{g}g=\sqrt{6\pi\rho}\left(\kappa_{crit}-\kappa\right)\,,\eeqa
which  relates  the external gravitational field $g$ with matter
density $\rho$.

Thus, we can readily see from \eq{g} that the slab is attractive
if $\kappa<\kappa_{crit}$, and  the whole space-time is trapped
between two singularities: the inner one discussed in the previous
section and the outer one at $z=1/3g$.

If $\kappa\leq0$, the range of the $z$-coordinate is
$-(\pi/2-u_\kappa)/\sqrt{6\pi\rho}<z<1/3g$, where $u_\kappa$
($0<u_\kappa<\pi/2$) is given implicitly in terms of $\kappa$
through \eq{k}. Inside the slab, the pressure is always positive,
and it diverges deep below at the inner singularity (see
Fig.\ref{Gpneg}). Its depth  is \beq
d=(\pi/2-u_\kappa)/\sqrt{6\pi\rho}\,. \eeq By using \eq{k}, we can
write $\kappa$ in terms of $d$ \beqa \lab{d} \kappa =-\frac{3 }{7
}\  \frac{{\cos^{7/3} (\sqrt{6\pi\rho}\,d)}}{\sin
(\sqrt{6\pi\rho}\,d)}\,\,\, _2F_1\!\Bigl(
1,\frac{2}{3};\frac{13}{6};\cos^2(\sqrt{6\pi\rho}\,d) \Bigr)\,.
\eeqa From \eq{g}, \eq{kapa} and \eq{d}, we can write the external
gravitational field $g$ in terms of the matter density $\rho$ and
the depth of the  slab $d$\beqa
g=\sqrt{6\pi\rho}\left(\frac{\sqrt{\pi}\;\Gamma(7/6)}{\Gamma(2/3)}+\frac{3
}{7 }\ \frac{{\cos^{7/3} (\sqrt{6\pi\rho}\,d)}}{\sin
(\sqrt{6\pi\rho}\,d)}\,\,\, _2F_1\!\Bigl(
1,\frac{2}{3};\frac{13}{6};\cos^2(\sqrt{6\pi\rho}\,d)
\Bigr)\right)\,.\eeqa

If $0<\kappa<\kappa_{crit}$,  the $z$-range is
$-\sqrt{\pi/24\rho}<z<1/3g$ and inside the slab, the pressure is
always finite, but it is negative deep below and $p=-\rho$ at the
inner singularity (see Fig.\ref{Gp}(c)).

Therefore, when $\kappa<\kappa_{crit}$, we see how the attraction
of the distant matter lying below $z=0$ shrinks the \st\ in such a
way that it finishes high above at the outer empty singular
boundary (at $z=1/3g$).

It can readily be seen from \eq{g} that, if
$\kappa>\kappa_{crit}$, $g$ is negative and  the  slab turns out
to be repulsive. In this case, the space-time is semi-infinite and
asymptotically flat when $z\rightarrow\infty$ (see \cite{gs}).
Inside the slab, the pressure is always finite and negative,  and
$p=-\rho$ at the inner singularity (see Fig.\ref{Gp}(a)).

If $\kappa=\kappa_{crit}$ it is {\em gravitationally neutral}, and
the exterior is one half of Minkowski's space-time.

Two remarks are in order. First, notice that the maximum depth
that a slab with constant density $\rho$ can reach is
$\sqrt{\pi/24\rho}$, being the counterpart of the well-known bound
$M<4R/9$ ($R<1/\sqrt{3\pi\rho}$) which holds for spherical
symmetry.

On the other hand, notice that the derivative of
$g_{xx}(z)=g_{yy}(z)$ has a discontinuity at the surface ($z=0$)
as it occurs with $\pd rg_{rr}(r)$ in the Schwarzschild's case
\cite{schw}, since
 \beqa  \pd
zg_{xx}(0)|_{interior}=\pd
zg_{yy}(0)|_{interior}=-\frac{4\sqrt{6\pi\rho}}{3}
\cos^{\frac{1}{3}} (\sqrt{6\pi\rho}\,z)\sin
(\sqrt{6\pi\rho}\,z)|_{z=0}=0\,,\eeqa but \beqa  \pd
zg_{xx}(0)|_{exterior}=\pd zg_{yy}(0)|_{exterior}=-4g
(1-3gz)^{1/3}|_{z=0}=-4g\,.\eeqa

\subsection{Matching two slabs}

Now we consider two incompressible fluids joined at $z=0$ where
the pressure vanishes, the lower one having a density $\rho$ and
the upper having a density $\rho'$. Thus, the lower solution is
given by \eq{met3}. By means of the transformation
$z\rightarrow-z$, $\rho\rightarrow\rho'$ and
$\kappa\rightarrow\kappa'$ we  get the upper one \beqa
\lab{met4}ds^2=  - G_{\kappa'}(\pi/2-\sqrt{6\pi\rho'}\,z)^{2}\,
dt^2 + \,\cos^{\frac{4}{3}}
(\sqrt{6\pi\rho'}\,z)\,\left(dx^2+dy^2\right) + dz^2,\nn
-\infty<t<\infty,\quad-\infty<x<\infty,\quad-\infty<y<\infty,\quad 0\leq z<\sqrt{\pi/24\rho'}\,. \eeqa%
From \eq{met3} and \eq{met4}, we can readily see that $g_{tt}(z)$,
$g_{xx}(z)$ and $\pd zg_{xx}(z)$ are continuous at $z=0$.
Furthermore, from \eq{G'2} we see that the continuity of  $\pd
zg_{tt}$ requires \beqa
\lab{G'4}\sqrt{\rho}\left(\kappa_{crit}-\kappa\right)=-\sqrt{\rho'}\left(\kappa_{crit}-\kappa'\right)\,.
 \eeqa
Thus, if one solution has a $\kappa$ greater than $\kappa_{crit}$,
the other must have it smaller than $\kappa_{crit}$. Therefore,
the joining is only possible between an attractive solution and a
repulsive one, or between two neutral ones.

It is easy to see that we can also insert  a slice of arbitrary
thickness of the vacuum solution \eq{taub} between them, obtaining
a full relativistic plane ``gravitational capacitor". For example,
we can trap a slice of Minkowski's \st\, between two solutions
with $\kappa=\kappa_{crit}$.

\section{Concluding remarks }

We have done a detailed study of  the exact solution of Einstein's
equations corresponding to a static and plane symmetric
distribution of  matter with constant  positive density. By
matching this internal solution to the vacuum one,  we showed that
different situations arise depending on the value of a parameter
$\kappa$. These  simple space-times turn out to present some
somehow astonishing properties.

For $\kappa<\kappa_{crit}$, the attraction of the  distant matter
shrinks the \st\ in such a way that it finishes  high above at an
empty singular boundary as  pointed out  in \cite{gs}. This
space-time also finishes down below at another singularity.

For $\kappa<0$, we have explicitly computed the external
gravitational field $g$ in terms of the density $\rho$ and the
sickness $d$ of the slab.

For $\kappa>\kappa_{crit}$,  negative but finite pressure
dominates the attraction of the matter and the slab turns out to
be repulsive.

We showed that the maximal sickness that these slabs can have is
$\sqrt{\pi/24\rho}$.

We have also discussed  matching an attractive slab to a repulsive
one, and two neutral ones. We also comment on how to assemble
relativistic gravitational capacitors consisting of a slice of
vacuum trapped between two of such slabs.

\section*{References}

\end{document}